\documentclass[12pt]{iopart}

\usepackage{iopams}

\usepackage{graphicx}
\usepackage{theorem}
\usepackage{hyperref}

\theoremstyle{plain}
\newtheorem{Thm}{Theorem}
\newtheorem{Lemma}[Thm]{Lemma}

\newtheorem{Prop}[Thm]{Proposition}
\newtheorem{Fact}[Thm]{Fact}

\newtheorem{Def}[Thm]{Definition}

\theorembodyfont{\rmfamily}
\newtheorem{Proof}{Proof}

\begin{document}

\title{Solutions to the ultradiscrete KdV equation expressed as the maximum of a quadratic function}
\author{Yoichi Nakata}
\address{Graduate School of Mathematical Sciences, the University of Tokyo, 3-8-1 Komaba, Meguro-ku, 153-8914 Tokyo, Japan}
\ead{ynakata@ms.u-tokyo.ac.jp}
\begin{abstract}
We propose the functions defined by the maximum of a discrete quadratic form and satisfying the ultradiscrete KdV equation. These functions includes not only soliton solutions but also pseudo-periodic solutions. In the proof, we employ some facts of discrete convex analysis.
\end{abstract}
\vspace{2pc}

\section{Introduction}
The ultradiscrete KdV equation
\begin{equation}\label{ukdv}
	T^{t+2}_{j+1} + T^t_j = \max \big( T^{t+2}_j + T^t_{j+1} - 1, T^{t+1}_j + T^{t+1}_{j+1} \big),
\end{equation}
is obtained from the bilinear form of the discrete KdV equation \cite{Hirota1977-1} by a limiting procedure called ``ultradiscretization" \cite{TakahashiTokihiroMatsukidairaSatsuma}. By setting
\begin{equation}\label{defu}
	U^t_j = T^{t+1}_j + T^t_{j+1} - T^{t+1}_{j+1} - T^t_j,
\end{equation}
this equation is transformed into
\begin{equation}\label{evBBS}
	U^{t+1}_j = \min \Big( 1 - U^t_j, \sum_{j'=j_0}^{j-1} (U^t_{j'} - U^{t+1}_{j'}) \Big),
\end{equation}
where $j_0$ is determined by the boundary conditions. This equation is known as the time evolution rule of the Box and Ball System (BBS) \cite{TakahashiSatsuma}, which is a cellular automaton with soliton like behavior in spite of the simple time evolution rules. The boundary conditions which are actively studied are $U^t_j = 0$ for $|j| \gg 1$ (infinite systems) and $U^t_{j+L} = U^t_j$ for some $L \in \mathbb{Z}_{>0}$ (periodic systems). We may choose $j_0$ as $j_0 = -\infty$ for infinite systems and can also set $j_0$ for periodic systems under some conditions.

These systems have good mathematical structures as well as continuous and discrete ones. For infinite systems, in previous papers \cite{Nakata2009}, \cite{Nakata2010}, we proposed a recursive representation which corresponds to the notion of vertex operators. As an analogue of determinant-type solutions, the ultradiscretization of signature-free determinants (called ``Permanents") is discussed in \cite{TakahashiHirota} and the relationship between this type of solution and ultradiscrete soliton equations is discussed in \cite{NagaiTakahashi2010} and \cite{NagaiTakahashi2011}. Approaches revealing combinatorial properties of the solutions are presented in \cite{TakagakiKamioka}, \cite{NoumiYamada} and \cite{Nakata2011}, by expressing them as maximum (minimum) weight flows of a planar graph.

For periodic systems, algebro-geometrical methods are considered to be most suitable and many topics are studied in the sense of this method. The initial value problem of periodic BBS (pBBS) is solved bypassing the analysis for ``discrete" (non-ultradiscrete) elliptic curve in \cite{KimijimaTokihiro}. An ultradiscrete closed method however is presented in \cite{InoueTakenawa}. A direct correspondence of these methods is presented in \cite{Iwao2009}.

A description of the dynamics of the BBS using the representation theory is presented in \cite{HatayamaHikamiInoueKunibaTakagi}. This approach is applicable to both infinite systems and periodic ones.

In this paper, we first consider a discrete quadratic function with a parameter and discuss the properties of this function, especially the values of the dependent variables where it attains its maximum. We prove that this maximum is a solution of the ultradiscrete KdV equation. We propose some examples of such solutions, which include well-known pseudo-periodic solutions and soliton solutions of the equation. Finally, we discuss the recursive representation we proposed earlier. The approach used here does not depend on algebro-geometric methods, but a few results of discrete convex analysis are used. 

\section{Discrete quadratic form}

Let $N$ be a natural number, ${\mathcal D}_i$ ($i=1, \ldots, N$) be a discrete interval $[a_i, b_i] \subset \mathbb{Z}$, a discrete semi-infinite inteval $(-\infty, b_i]$, $[a_i, \infty)$ or $\mathbb{Z}$ and ${\mathcal D} = {\mathcal D}_1 \times \ldots \times {\mathcal D}_N \subset \mathbb{Z}^N$. We consider a discrete quadratic function for $\mathbf{m} \in {\mathcal D}$ with parameters $\mathbf{z} \in \mathbb{R}^N$, defined as
\begin{eqnarray}\label{deff}
	f(\mathbf{z}; \mathbf{m}) = \frac{1}{2} {}^t \!\mathbf{m} A \mathbf{m} + {}^t \!\mathbf{z}  \mathbf{m},
\end{eqnarray}
where the matrix $A \in {\mathrm{mat}}(\mathbb{R}, N)$ is given by
\begin{equation}\label{defA}
	(A)_{i, k} = \cases{- L_i + 2\sum_{l=1}^{i-1} \Omega_l + 2 (N-i) \Omega_i  & ($i=k$) \\ -2\Omega_i & ($i<k$) \\ -2\Omega_k & ($i>k$)}
\end{equation}
and parameters $\Omega_i, L_i \in \mathbb{R}$ satisfying the relations:
\begin{eqnarray}
	&&1 \le \Omega_1 \le \Omega_2 \le \ldots \le \Omega_N \label{condomega} \\
	&&2\sum_{l=1}^{i-1} \Omega_l + 2 (N-i+1) \Omega_i < L_i \quad (i = 1, \ldots, N). \label{condL}
\end{eqnarray}

By introducing a matrix $M \in {\mathrm{mat}}(\mathbb{R}, N)$ given by
\begin{equation}
	(M)_{i, k} = \delta_{i, k} - \delta_{i+1, k}
\end{equation}
and ${\boldsymbol \mu} \in \mathbb{Z}^N$ expressed as
\begin{equation}
	\mathbf{m} = M {\boldsymbol \mu} \label{mutom},
\end{equation}
the function $f$ is transformed into:
\begin{equation}\label{deftau2}
	g(\mathbf{z}; {\boldsymbol \mu}) = \frac{1}{2} {}^t \! {\boldsymbol \mu} B {\boldsymbol \mu} + {}^t \!\mathbf{z} M {\boldsymbol \mu},
\end{equation}
where $B ={}^t \! M A M$, is found to be
\begin{equation}
	(B)_{i, k} = \cases{ (A)_{1, 1} & ($i=k=1$) \\ (A)_{i-1, i-1} + (A)_{i, i} + 4 \Omega_{i-1} & ($i=k \ge 2$) \\  - (A)_{i-1, i-1} - 2 \Omega_{i-1} & ($i-k = 1$) \\ - (A)_{i, i} - 2 \Omega_i & ($i-k = -1$) \\ 0 & ($|i-k| \ge 2$) }.
\end{equation}
It should be noted that these two quadratic functions are equivalent because $M^{-1}$ is expressed as
\begin{equation}
	(M^{-1})_{i, k} = \cases{ 1 & ($i\ge k$) \\ 0 & ($i<k$)}.
\end{equation}

\begin{Prop}\label{negative}
$B$ is negative definite.
\end{Prop}

\begin{Proof}
We separate the matrix $B$ into $B = B_1 + B_2$, with $B_1$ and $B_2$ expressed as 
\begin{eqnarray}
	(B_1)_{i, k} = \cases{ (A)_{1, 1} + 2 \Omega_1 & ($i=k=1$) \\ (A)_{i-1, i-1} + 2 \Omega_{i-1} + (A)_{i, i} + 2 \Omega_i & ($i=k \ge 2$) \\  - (A)_{i-1, i-1} - 2 \Omega_{i-1} & ($i-k = 1$) \\ - (A)_{i, i} - 2 \Omega_i & ($i-k = -1$) \\ 0 & ($|i-k| \ge 2$) } \\
	B_2 = 2 \,{\mathrm{diag}}( -\Omega_1, \Omega_1 - \Omega_2, \ldots, \Omega_{N-1} - \Omega_N ).
\end{eqnarray}
Here, $B_1$ can be transformed into
\begin{equation}
	{}^t\! (M^{-1}) B_1 M^{-1} = {\mathrm{diag}}( (A)_{1, 1} + 2 \Omega_1, \ldots, (A)_{N, N} + 2 \Omega_N ),
\end{equation}
which is negative by condition (\ref{condL}) and $B_2$ is at least non-positive by (\ref{condomega}). \hfill \opensquare
\end{Proof}

Due to this proposition, the quadratic function $f(\mathbf{z}; \mathbf{m}) = g(\mathbf{z}; {\boldsymbol \mu})$ always has a maximum and the set on which it elements attains its maximum is finite.

\begin{Fact}\label{supermodular}
The quadratic form $\frac{1}{2} {}^t \!{\boldsymbol \mu} B {\boldsymbol \mu}$ satisfies the relation:
\begin{equation}
	 \frac{1}{2} {}^t \!{\boldsymbol \mu} B {\boldsymbol \mu} + \frac{1}{2} {}^t \!{\boldsymbol \mu} B {\boldsymbol \mu} \le  \frac{1}{2} {}^t \!({\boldsymbol \mu} \wedge {\boldsymbol \mu}') B ({\boldsymbol \mu} \wedge {\boldsymbol \mu}') +  \frac{1}{2} {}^t \!({\boldsymbol \mu} \vee {\boldsymbol \mu}') B ({\boldsymbol \mu} \vee {\boldsymbol \mu}'),
\end{equation}
where $\wedge$ and $\vee$ means
\begin{equation}
	({\boldsymbol \mu} \wedge {\boldsymbol \mu}')_i := \min ( \mu_i, \mu'_i )  \qquad ({\boldsymbol \mu} \vee {\boldsymbol \mu}')_i := \max ( \mu_i, \mu'_i ).
\end{equation}
\end{Fact}

This fact is a famous result of convex analysis and a proof is presented, for example, in \cite{Murota2003}.

\begin{Def}
We introduce an ordering for ${\boldsymbol \mu}$ by
\begin{equation}\label{deforder}
	{\boldsymbol \mu} \ge {\boldsymbol \mu}'    \Longleftrightarrow \mu_i \ge \mu'_i \ \ \mbox{for} \ 1 \le i \le N.
\end{equation}
\end{Def}

\begin{Prop}\label{Lconvmaximalset}
The set of ${\boldsymbol \mu}$ which realizes the maximum of (\ref{deftau2}) has a unique maximum element with respect to this ordering.
\end{Prop}

\begin{Proof}
We assume the existence of two different local-maximal elements ${\boldsymbol \mu}$ and ${\boldsymbol \mu}'$ and denote ${\boldsymbol \mu}'' = {\boldsymbol \mu} \wedge {\boldsymbol \mu}'$ and ${\boldsymbol \mu}''' = {\boldsymbol \mu} \vee {\boldsymbol \mu}'$.  Due to Fact \ref{supermodular} and identity: ${\boldsymbol \mu} + {\boldsymbol \mu}' = {\boldsymbol \mu}'' + {\boldsymbol \mu}'''$, we obtain
\begin{equation}
		\frac{1}{2} {}^t \! {\boldsymbol \mu} B {\boldsymbol \mu} + {}^t \!\mathbf{z} M {\boldsymbol \mu} + \frac{1}{2} {}^t \! {\boldsymbol \mu}' B {\boldsymbol \mu}' + {}^t \!\mathbf{z} M {\boldsymbol \mu}' \le \frac{1}{2} {}^t \! {\boldsymbol \mu}'' B {\boldsymbol \mu}'' + {}^t \!\mathbf{z} M {\boldsymbol \mu}'' + \frac{1}{2} {}^t \! {\boldsymbol \mu}''' B {\boldsymbol \mu}''' + {}^t \!\mathbf{z} M {\boldsymbol \mu}'''.
\end{equation}
However, this inequality is actually an equality because ${\boldsymbol \mu}$ and ${\boldsymbol \mu}'$ yield a maximum, i.e., ${\boldsymbol \mu}''$ and ${\boldsymbol \mu}'''$ also yield a maximum. This contradicts that ${\boldsymbol \mu}$ and ${\boldsymbol \mu}'$ are local-maximum because ${\boldsymbol \mu}''' > {\boldsymbol \mu}, {\boldsymbol \mu}'$. \hfill \opensquare
\end{Proof}

\begin{Def}
For each $t, j \in \mathbb{Z}$, let
\begin{eqnarray}
	\mathbf{z}^t_j = t \mathbf{\Omega} - j \mathbf{1} + \mathbf{C} \\
	\mathbf{\Omega} = {}^t ( \Omega_1, \ldots, \Omega_N ) \\
	\mathbf{C} = {}^t ( C_1, \ldots, C_N )\\
	\mathbf{1} = {}^t ( 1, \ldots, 1 ).
\end{eqnarray}
We define ${\boldsymbol \mu}^{(N), t}_j$ as ${\boldsymbol \mu}$ which yields the maximum of $g(\mathbf{z}^t_j; {\boldsymbol \mu})$ and which is the maximal element for the ordering (\ref{deforder}) for such ${\boldsymbol \mu}$. We also denote $\mathbf{m}^{(N), t}_j := M {\boldsymbol \mu}^{(N), t}_j$ and $T^{(N), t}_j$ as the maximum value of $g(\mathbf{z}^t_j; {\boldsymbol \mu}) = f(\mathbf{z}^t_j; \mathbf{m})$ , i.e., 
\begin{equation}\label{deftau}
	T^{(N), t}_j = \max_{\mathbf{m} \in \mathbb{Z}^N} f(\mathbf{z}^t_j; \mathbf{m}) = f(\mathbf{z}^t_j; \mathbf{m}^{(N), t}_j) = g(\mathbf{z}^t_j; {\boldsymbol \mu}^{(N), t}_j)
\end{equation}
We simply denote $T^t_j$ and $\mathbf{m}^t_j$ when we do not need to consider the value of $N$.
\end{Def}

Let $m_N$ be the $N$-th component of $\mathbf{m}$ and $\tilde{\mathbf{m}}$ be a subvector consisting of the first $N-1$ components of $\mathbf{m}$, $f(\mathbf{z}^t_j; \mathrm{m})$ is transformed into
\begin{equation}\label{vertexop0}
	 f(\mathbf{z}^t_j; \mathbf{m}) = \frac{1}{2} \Big( - L_N + 2 \sum_{i=1}^{N-1} \Omega_i \Big) m^2_N + z_N m_N + \tilde{f}(\tilde{\mathbf{z}}^{t-2 m_N}_j; \tilde{\mathbf{m}}),
\end{equation}
where $\tilde{f}$ is the discrete quadratic function for $\tilde{\mathbf{m}}$ written in 
\begin{equation}
	\tilde{f}(\tilde{\mathbf{z}}; \tilde{\mathbf{m}}) = \frac{1}{2} {}^t \!\tilde{\mathbf{m}} \tilde{A} \tilde{\mathbf{m}} + {}^t \!\tilde{\mathbf{z}}  \tilde{\mathbf{m}},
\end{equation}
matrix $\tilde{A}$ is a submatrix of $A$ consisting of the first $N-1$ rows and columns and $\tilde{\mathbf{z}}^{t - 2 m_N}_j \in \mathbb{R}^{N-1}$ is a subvector of  $\mathbf{z}^{t-2 m_N}_j$ consisting of the first $N-1$ rows. In other words, $\tilde{A}$ is obtained by replacing $L_i \to L_i - 2 \Omega_i$ ($i=1, \ldots, N-1$) in the definition (\ref{defA}) for $N-1$. It should be noted that the condition (\ref{condL}) is also satisfied for $\tilde{f}$ when all parameters $L_i$ and $\Omega_i$ are fixed. Therefore, $\tilde{f}$ has the maximum value. By denoting this maximum for $\tilde{\mathbf{z}}^t_j$ as $\tilde{T}^{(N-1), t}_j$, we obtain the recursive form
\begin{equation}\label{vertexop}
	T^{(N), t}_j = \max_{m_N \in \mathbb{Z}} \left( \frac{1}{2} \Big( - L_N + 2 \sum_{i=1}^{N-1} \Omega_i \Big) m^2_N + z_N m_N + \tilde{T}^{(N-1), t - 2 m_N}_j \right).
\end{equation}
Because of the above discussion, the first $N-1$ components of $\mathbf{m}^{(N), t}_j$ are equal to $\tilde{\mathbf{m}} \in \mathbb{Z}^{N-1}$ which yields the maximum of $\tilde{f}(\tilde{\mathbf{z}}^{t - 2 m_N}_j; \tilde{\mathbf{m}})$, where $m_N$ is the $N$-th component of $\mathbf{m}^{(N), t}_j$. We also define $\tilde{f} \equiv 0$ for $N=1$ (i.e. $T^{(0), t}_j \equiv 0$), for consistency.

\section{Behaviour of maximizing vectors}

To prove that $T^{(N), t}_j$ solves the ultradiscrete KdV equation, we study the behaviour of $\mathbf{m}^{(N), t}_j$.

\begin{Lemma}\label{relmu}
${\boldsymbol \mu}^t_j$ satisfies the relation:
\begin{equation}
	{\boldsymbol \mu}^t_j < {\boldsymbol \mu}^t_{j-1} < {\boldsymbol \mu}^{t+1}_j.
\end{equation}
\end{Lemma}

\begin{Proof}
We denote $\mathbf{z}^t_j = \mathbf{z}$, ${\boldsymbol \mu}^t_j = {\boldsymbol \mu}$, ${\boldsymbol \mu}^t_{j-1} = {\boldsymbol \mu}'$, ${\boldsymbol \mu}'' = {\boldsymbol \mu} \wedge {\boldsymbol \mu}'$ and ${\boldsymbol \mu}''' = {\boldsymbol \mu} \vee {\boldsymbol \mu}'$. Due to Fact \ref{supermodular} and $\mathbf{z}^t_{j-1} = \mathbf{z}^t_j+\mathbf{1}$, we obtain
\begin{eqnarray}
	\frac{1}{2} {}^t \! {\boldsymbol \mu} B {\boldsymbol \mu} + {}^t \!\mathbf{z} M {\boldsymbol \mu} + \frac{1}{2} {}^t \! {\boldsymbol \mu}' B {\boldsymbol \mu}' + {}^t \!(\mathbf{z} + \mathbf{1} ) M {\boldsymbol \mu}' \nonumber\\ 
\le \frac{1}{2} {}^t \! {\boldsymbol \mu}'' B {\boldsymbol \mu}'' + {}^t \!\mathbf{z} M {\boldsymbol \mu}'' + \frac{1}{2} {}^t \! {\boldsymbol \mu}''' B {\boldsymbol \mu}''' + {}^t \!(\mathbf{z} + \mathbf{1}) M {\boldsymbol \mu}''' - {}^t \!\mathbf{1} M ({\boldsymbol \mu}''' - {\boldsymbol \mu}').
\end{eqnarray}
Here, one has ${}^t \!\mathbf{1} M ({\boldsymbol \mu}''' - {\boldsymbol \mu}') \ge 0$ because of the definition of ${\boldsymbol \mu}'''$. Thus, we obtain
\begin{eqnarray}
	\frac{1}{2} {}^t \! {\boldsymbol \mu} B {\boldsymbol \mu} + {}^t \!\mathbf{z} M {\boldsymbol \mu} + \frac{1}{2} {}^t \! {\boldsymbol \mu}' B {\boldsymbol \mu}' + {}^t \!(\mathbf{z} + \mathbf{1} ) M {\boldsymbol \mu}' \nonumber\\
	\le \frac{1}{2} {}^t \! {\boldsymbol \mu}'' B {\boldsymbol \mu}'' + {}^t \!\mathbf{z} M {\boldsymbol \mu}'' + \frac{1}{2} {}^t \! {\boldsymbol \mu}''' B {\boldsymbol \mu}''' + {}^t \!(\mathbf{z} + \mathbf{1}) M {\boldsymbol \mu}'''.
\end{eqnarray}
By the same discussion as in the proof of Proposition \ref{Lconvmaximalset}, ${\boldsymbol \mu}''$ and ${\boldsymbol \mu}'''$ must yield a maximum for $(t, j)$ and $(t, j-1)$ respectively. By virtue of the maximality of ${\boldsymbol \mu}'$ and the definition of ${\boldsymbol \mu}'''$, we obtain ${\boldsymbol \mu}^t_{j-1} = {\boldsymbol \mu}' = {\boldsymbol \mu}''' > {\boldsymbol \mu} = {\boldsymbol \mu}^t_j$.

The proof for ${\boldsymbol \mu}^t_j < {\boldsymbol \mu}^{t+1}_{j+1}$ is completely the same. \hfill \opensquare
\end{Proof}

\begin{Thm}\label{relt}
$\mathbf{m}^{t-2}_j$ is the same as $\mathbf{m}^t_j$ or expressed as $\mathbf{m}^t_j = \mathbf{m}^{t-2}_j + \mathbf{e}_i \in {\mathcal D}$ for some $1 \le i \le N$. Here, $\mathbf{e}_i$ is the $i$-th canonical basis vector.
\end{Thm}

Before giving the proof, we prepare a lemma.

\begin{Lemma}\label{convex}
Under Theorem \ref{relt}, one has
\begin{equation}
	T^{(N), t+2}_j + T^{(N), t-2}_j \le 2 T^{(N), t}_j + 2 \Omega_N.
\end{equation}
\end{Lemma}

\begin{Proof}
By denoting $\mathbf{m}^{(N), t+2}_j = \mathbf{m}$, $\mathbf{m}^{(N), t-2}_j = \mathbf{m}'$ and $\mathbf{z}^t_j = \mathbf{z}$, one has
\begin{equation}\label{convex-mid1}
	T^{(N), t+2}_j + T^{(N), t-2}_j = \frac{1}{2} {}^t \mathbf{m} A \mathbf{m} + {}^t (\mathbf{z} + 2 \mathbf{\Omega}) \mathbf{m} + \frac{1}{2} {}^t \mathbf{m}' A \mathbf{m}' + {}^t (\mathbf{z} - 2 \mathbf{\Omega}) \mathbf{m}'.
\end{equation}
Due to Theorem \ref{relt}, $\mathbf{m} - \mathbf{m}'$ is equal to $\mathbf{0}$, $\mathbf{e}_i$ or $\mathbf{e}_i + \mathbf{e}_k$. In the case of $\mathbf{m} = \mathbf{m}'$, $\mathbf{m}^t_j$ is also equal to $\mathbf{m}$. Then, we obtain
\begin{eqnarray}
	T^{(N), t+2}_j + T^{(N), t-2}_j = \frac{1}{2} {}^t \mathbf{m} A \mathbf{m} + {}^t \mathbf{z} \mathbf{m} + \frac{1}{2} {}^t \mathbf{m} A \mathbf{m} + {}^t \mathbf{z} \mathbf{m} \nonumber\\
	= 2  T^{(N), t}_j < 2  T^{(N), t}_j + 2 \Omega_N.
\end{eqnarray}
In the case of $\mathbf{m} = \mathbf{m}' + \mathbf{e}_i$, we obtain
\begin{eqnarray}
	T^{(N), t+2}_j + T^{(N), t-2}_j \nonumber\\
        = \frac{1}{2} {}^t (\mathbf{m}' + \mathbf{e}_i) A (\mathbf{m}' + \mathbf{e}_i) + {}^t \mathbf{z} (\mathbf{m}' + \mathbf{e}_i ) + \frac{1}{2} {}^t \mathbf{m}' A \mathbf{m}' + {}^t \mathbf{z} \mathbf{m}' + 2 \Omega_i  \nonumber\\
	\le 2  T^{(N), t}_j + 2 \Omega_i \le 2  T^{(N), t}_j + 2 \Omega_N.
\end{eqnarray}
In the case of $\mathbf{m} = \mathbf{m}' + \mathbf{e}_i + \mathbf{e}_k$, due to identity:
\begin{eqnarray}
	\frac{1}{2} {}^t \!(\mathbf{m} + \mathbf{e}_i) A (\mathbf{m} + \mathbf{e}_i) + \frac{1}{2} {}^t \!(\mathbf{m} + \mathbf{e}_k) A (\mathbf{m} + \mathbf{e}_k) \nonumber\\ 
        = \frac{1}{2} {}^t \!(\mathbf{m} + \mathbf{e}_i + \mathbf{e}_k) A (\mathbf{m} + \mathbf{e}_i + \mathbf{e}_k) + \frac{1}{2} {}^t \!\mathbf{m} A \mathbf{m} - (A)_{i, k} \label{sqidentity},
\end{eqnarray}
we obtain
\begin{eqnarray}
	T^{(N), t+2}_j + T^{(N), t-2}_j = \frac{1}{2} {}^t (\mathbf{m}' + \mathbf{e}_i) A (\mathbf{m}' + \mathbf{e}_i) + {}^t \mathbf{z} (\mathbf{m}' + \mathbf{e}_i ) \nonumber\\ 
+ \frac{1}{2} {}^t (\mathbf{m}' + \mathbf{e}_k )  A ( \mathbf{m}' + \mathbf{e}_k ) + {}^t \mathbf{z} ( \mathbf{m}' + \mathbf{e}_k ) + 2 \Omega_i + 2 \Omega_k + (A)_{i, k}.
\end{eqnarray}
Here, by the definition of $(A)_{i, k}$, one has
\begin{equation}
	2 \Omega_i + 2 \Omega_k + (A)_{i, k} = \cases{ -L_i + 2\sum_{l=1}^{i-1} \Omega_l + 2 (N-i) \Omega_i + 2 \Omega_i & ($i=k$) \\ 2 \Omega_{\max(i, k)} & ($i \neq k$) }
\end{equation}
and this value is less than $2 \Omega_N$, in both cases, by virtue of (\ref{condomega}) and (\ref{condL}). \hfill \opensquare
\end{Proof}

Now, let us prove Theorem \ref{relt}.

\begin{Proof}

We employ the inductive method for $N$. It is clear that $(\mathbf{m}^{t-2}_j)_N \le (\mathbf{m}^t_j)_N$ because ${\boldsymbol \mu}^{t-2}_j < {\boldsymbol \mu}^t_j$. By denoting $(\mathbf{m}^t_j)_N = m_N$ and $(\mathbf{m}^{t-2}_j)_N = m'_N$ and employing (\ref{vertexop}), $T^{(N), t}_j$ and $T^{(N), t-2}_j$ are expressed as
\begin{eqnarray}
	T^{(N), t}_j = \frac{1}{2} \Big( - L_N + 2 \sum_{i=1}^{N-1} \Omega_i \Big) m^2_N + z_N m_N + \tilde{T}^{(N-1), t-2 m_N}_j \\
	\!\!\!\!\!\!\!\!\!\!\!\! T^{(N), t-2}_j = \frac{1}{2} \Big( - L_N + 2 \sum_{i=1}^{N-1} \Omega_i \Big) m'^2_N + \Big(z_N - 2 \Omega_N \Big) m'_N + \tilde{T}
^{(N-1), t-2 m'_N - 2}_j
\end{eqnarray}
and by the maximality for $T^{(N), t}_j$ and $T^{(N), t-2}_j$, one has
\begin{eqnarray}
	T^{(N), t}_j \ge \frac{1}{2} \Big( - L_N + 2 \sum_{i=1}^{N-1} \Omega_i \Big) (m_N - 1)^2  \nonumber\\
	\qquad \qquad \qquad + z_N (m_N - 1)+ \tilde{T}^{(N-1), t-2 m_N + 2}_j \\
	T^{(N), t-2}_j \ge \frac{1}{2} \Big( - L_N + 2 \sum_{i=1}^{N-1} \Omega_i \Big) (m'_N + 1)^2 \nonumber\\
	\qquad \qquad \qquad + \Big(z_N - 2 \Omega_N \Big) (m'_N + 1)  + \tilde{T}^{(N-1), t-2 m'_N - 4}_j.
\end{eqnarray}
Adding these equalities and inequalities, we obtain
\begin{eqnarray}\label{relt-mid1}
	0 \ge \Big( L_N - 2 \sum_{i=1}^{N-1} \Omega_i \Big) (m_N - m'_N - 1) - 2 \Omega_N \nonumber\\ 
	+ \tilde{T}^{(N-1), t-2 m_N + 2}_j + \tilde{T}^{(N-1), t-2 m'_N - 4}_j - \tilde{T}^{(N-1), t-2 m_N}_j - \tilde{T}^{(N-1), t-2 m'_N - 2}_j.
\end{eqnarray}
By assuming that this theorem holds for $N-1$, we can apply Lemma \ref{convex} repeatedly and obtain
\begin{eqnarray}
	\tilde{T}^{(N-1), t-2 m_N + 2}_j + \tilde{T}^{(N-1), t-2 m'_N - 4}_j - \tilde{T}^{(N-1), t-2 m_N}_j - \tilde{T}^{(N-1), t-2 m'_N - 2}_j  \nonumber\\
	\qquad \qquad \qquad \qquad \qquad \qquad \qquad \qquad \ge - 2 \Omega_N (m_N - m'_N - 2).
\end{eqnarray}
Therefore, (\ref{relt-mid1}) is transformed into
\begin{equation}
	0 \ge \Big( L_N - 2 \sum_{i=1}^N \Omega_i \Big) (m_N - m'_N - 1).
\end{equation}
For $N=1$, we also obtain the same result because $\tilde{T}^{(0), t}_j$ is equal to $0$. To satisfy this inequality, $m_N - m'_N$ must be less than $1$ because  of (\ref{condomega}) and (\ref{condL}).

Thus, we obtain $(\mathbf{m}^t_j)_N -1 \le (\mathbf{m}^{t-2}_j)_N \le (\mathbf{m}^t_j)_N$. If $(\mathbf{m}^t_j)_N = (\mathbf{m}^{t-2}_j)_N + 1$,  by the relation (\ref{vertexop}), the $i$-th components of $\mathbf{m}^t_j$ and $\mathbf{m}^{t-2}_j$ are the same for $i \le N-1$. If $(\mathbf{m}^t_j)_N = (\mathbf{m}^{t-2}_j)_N$, it is clear by the assumption of the induction. \hfill \opensquare
\end{Proof}

\begin{Lemma}\label{relj}
$\mathbf{m}^t_{j+1}$ is equal to $\mathbf{m}^t_j$ or express as $\mathbf{m}^t_j = \mathbf{m}^t_{j+1} + \mathbf{e}_i  \in {\mathcal D}$ for some $1 \le i \le N$.
\end{Lemma}

\begin{Proof}
Note that ${\boldsymbol \mu}^t_j = {\boldsymbol \mu}^{t-2}_j$ or
\begin{equation}
	{\boldsymbol \mu}^t_j = {\boldsymbol \mu}^{t-2}_j + \sum_{i=1}^{i_0} \mathbf{e}_i
\end{equation}
for some $i_0$ by virtue of Theorem \ref{relt}. Due to Proposition \ref{relmu}, ${\boldsymbol \mu}^t_{j+1}$ can be equal to ${\boldsymbol \mu}^t_j$, expressed as
\begin{equation}\label{relja}
	{\boldsymbol \mu}^t_{j+1} = {\boldsymbol \mu}^t_j - \sum_{i'=1}^{\tilde{i}} \mathbf{e}_{i'}
\end{equation}
for some $i_0 \ge \tilde{i}$ or
\begin{equation}\label{reljb}
	{\boldsymbol \mu}^t_{j+1} = {\boldsymbol \mu}^t_j - \sum_{i'=i_2+1}^{i_1} \mathbf{e}_{i'} -  \ldots - \sum_{i'=i_{2l}+1}^{i_{2l-1}} \mathbf{e}_{i'} - \sum_{i'=1}^{\tilde{i}} \mathbf{e}_{i'}
\end{equation}
for some $i_0 \ge i_1 > i_2 > \ldots > i_{2l-1} > i_{2l} > \tilde{i} $. 

We have to prove ${\boldsymbol \mu}^t_{j+1}$ cannot take the form (\ref{reljb}). We assume ${\boldsymbol \mu}^t_{j+1}$ takes the form: 
\begin{equation}\label{reljbp}
	{\boldsymbol \mu}^t_{j+1} = {\boldsymbol \mu}^t_j - \sum_{i'=i_2+1}^{i_1} \mathbf{e}_{i'} -  \sum_{i'=1}^{\tilde{i}} \mathbf{e}_{i'},
\end{equation}
which is the simplest case of (\ref{reljb}) and equivalent to $\mathbf{m}^t_{j+1} = \mathbf{m}^t_j - \mathbf{e}_{i_2} + \mathbf{e}_{i_1} - \mathbf{e}_{i'}$. Let
\begin{equation}
	{\boldsymbol \mu}' = {\boldsymbol \mu}^t_j - \sum_{i'=i_2+1}^{i_1} \mathbf{e}_{i'} \qquad \left( \mathbf{m}' = \mathbf{m}^t_j - \mathbf{e}_{i_2} + \mathbf{e}_{i_1}\right)
\end{equation}
and
\begin{equation}
	{\boldsymbol \mu}'' = {\boldsymbol \mu}^t_j - \sum_{i'=1}^{\tilde{i}} \mathbf{e}_{i'} \qquad \left( \mathbf{m}'' = \mathbf{m}^t_j - \mathbf{e}_{i'} \right).
\end{equation}
We also denote ${\boldsymbol \mu}^t_{j+1} = {\boldsymbol \mu}'''$ ($\mathbf{m}^t_{j+1} = \mathbf{m}'''$) and ${\boldsymbol \mu}^t_j = {\boldsymbol \mu}$ ($\mathbf{m}^t_j = \mathbf{m}$). It should be noted that $({\boldsymbol \mu})_{i_2} = ({\boldsymbol \mu}')_{i_2} = ({\boldsymbol \mu}'')_{i_2} = ({\boldsymbol \mu}''')_{i_2}$. We let $\mathbf{m}_1$ and $\mathbf{m}'_1$ be the vectors consisting of the components $i_2$ to $N$ of $\mathbf{m}$ and $\mathbf{m}'''$.  We also let $\mathbf{m}_2$ and $\mathbf{m}'_2$ be the vectors consisting of the components $1$ to $i_2-1$  of $\mathbf{m}$ and $\mathbf{m}'''$. In other words, $\mathbf{m} = {}^t\!({}^t\!\mathbf{m}_1, {}^t\!\mathbf{m}_2)$ and $\mathbf{m}''' = {}^t\!({}^t\!\mathbf{m}'_1, {}^t\!\mathbf{m}'_2)$. Then, $\mathbf{m}'$ and $\mathbf{m}''$ are also expressed as $\mathbf{m}' = {}^t\!({}^t\!\mathbf{m}'_1, {}^t\!\mathbf{m}_2)$ and $\mathbf{m}'' = {}^t\!({}^t\!\mathbf{m}_1, {}^t\!\mathbf{m}'_2)$.

By employing (\ref{vertexop0}) repeatedly, one can rewrite
\begin{eqnarray}
	f(\mathbf{z}^t_j; \mathbf{m}) = S(\mathbf{m}_1, t) - \mu_{i_2} j + \overline{f}({\overline{\mathbf{z}}^{t - \mu_{i_2}}_j}; \mathbf{m}_2) \\
	f(\mathbf{z}^t_j; \mathbf{m}') = S(\mathbf{m}'_1, t) - \mu_{i_2} j + \overline{f}({\overline{\mathbf{z}}^{t - \mu_{i_2}}_j}; \mathbf{m}_2).
\end{eqnarray}
Here, $S$ is a function which depends on $\mathbf{m}_1 \in \mathbb{Z}^{N-i_2}$ and $t$ but not on $j$, and $\overline{f}$ depends on $\overline{\mathbf{z}} \in \mathbb{R}^{i_2}$ and $\mathbf{m}_2 \in \mathbb{Z}^{i_2}$. The maximality of $\mathbf{m}$ leads to
\begin{equation}
	S(\mathbf{m}_1, t) - S(\mathbf{m}'_1, t) \ge 0.
\end{equation}
However, by the same discussion for $j+1$, one has
\begin{eqnarray}
	f(\mathbf{z}^t_{j+1}; \mathbf{m}'') = S(\mathbf{m}_1, t) - \mu_{i_2} (j+1) + \overline{f}({\overline{\mathbf{z}}^{t - \mu_{i_2}}_{j+1}}; \mathbf{m}'_2) \\
	f(\mathbf{z}^t_{j+1}; \mathbf{m}''') = S(\mathbf{m}'_1, t) - \mu_{i_2} (j+1) + \overline{f}({\overline{\mathbf{z}}^{t - \mu_{i_2}}_{j+1}}; \mathbf{m}'_2).
\end{eqnarray}
Since $\mathbf{m}''$ attains the maximum value of $f(\mathbf{z}^t_{j+1}; \cdots)$ and the maximum element for the ordering (\ref{deforder}), we obtain
\begin{equation}
	S(\mathbf{m}_1; t) - S(\mathbf{m}'_1; t) < 0,
\end{equation}
which is a contradiction. It is clear that we can extend this proof to the general case of (\ref{reljb}). \opensquare
\end{Proof}

By employing all these lemmas and enumerating all possibilities, we finally obtain the important theorem:
\begin{Thm}\label{Thm1}
Vectors $\mathbf{m}^{t+2}_j$, $\mathbf{m}^{t+2}_{j+1}$, $\mathbf{m}^{t+1}_j$, $\mathbf{m}^{t+1}_{j+1}$, $\mathbf{m}^t_j$ and $\mathbf{m}^t_{j+1}$ are expressed as
\begin{eqnarray}
	(\mathbf{m}^{t+2}_j, \mathbf{m}^{t+2}_{j+1}, \mathbf{m}^{t+1}_j, \mathbf{m}^{t+1}_{j+1}, \mathbf{m}^t_j, \mathbf{m}^t_{j+1})  \nonumber\\
	\qquad \qquad \qquad = (\mathbf{m}^t_{j+1}, \mathbf{m}^t_{j+1}, \mathbf{m}^t_{j+1}, \mathbf{m}^t_{j+1}, \mathbf{m}^t_{j+1}, \mathbf{m}^t_{j+1}) + \Delta,
\end{eqnarray}
where $\Delta$ is one of the following:
\begin{eqnarray}
	(\mathbf{0}, \mathbf{0}, \mathbf{0}, \mathbf{0}, \mathbf{0}, \mathbf{0}) \\
	(\mathbf{e}_i, \mathbf{0}, \mathbf{0}, \mathbf{0}, \mathbf{0}, \mathbf{0}) \\
	(\mathbf{e}_i, \mathbf{e}_i, \mathbf{d}, \mathbf{d}, \mathbf{0}, \mathbf{0}) \\
	(\mathbf{e}_i, \mathbf{e}_i, \mathbf{d} + \mathbf{e}_k, \mathbf{d}, \mathbf{0}, \mathbf{0}) \\
	(\mathbf{e}_i, \mathbf{e}_i, \mathbf{d} + \mathbf{e}_k, \mathbf{d} + \mathbf{e}_k, \mathbf{0}, \mathbf{0}) \\
 	(\mathbf{e}_i, \mathbf{e}_i, \mathbf{e}_i, \mathbf{e}_i, \mathbf{e}_i, \mathbf{0}) \\
 	(2 \mathbf{e}_i, \mathbf{e}_i, \mathbf{e}_i, \mathbf{e}_i, \mathbf{e}_i, \mathbf{0}) \\
 	(\mathbf{e}_i + \mathbf{e}_l, \mathbf{e}_i, \mathbf{e}_i, \mathbf{e}_i, \mathbf{e}_i, \mathbf{0}) \\
	(\mathbf{e}_i + \mathbf{e}_k, \mathbf{e}_i, \mathbf{d} + \mathbf{e}_{k'}, \mathbf{d} + \mathbf{e}_{k'}, \mathbf{e}_k, \mathbf{0}).
\end{eqnarray}
Here, $i$ and $k$ satisfy $i>k$ and $\mathbf{d}$ is expressed as
\begin{eqnarray}
	\mathbf{d} = \mathbf{e}_{i_1} - \mathbf{e}_{i_2} + \ldots + \mathbf{e}_{i_{2l-1}} - \mathbf{e}_{i_{2l}}
\end{eqnarray}
for some $i > i_1 > i_2 > \ldots > i_{2l-1} > i_{2l} > k' \ge k \ge 0$. 
\end{Thm}

\section{The ultradiscrete KdV equation}

\begin{Thm}\label{mainThm}
The function $T^t_j$ satisfies the ultradiscrete KdV equation:
\begin{equation}
	T^{t+2}_{j+1} + T^t_j = \max ( T^{t+2}_j + T^t_{j+1} - 1, T^{t+1}_j + T^{t+1}_{j+1} ).
\end{equation}
\end{Thm}

\begin{Proof}
We denote $\mathbf{m}^t_{j+1} = \mathbf{m}$ and $\mathbf{z}^t_j = \mathbf{z}$. It is sufficient to prove each case of $\Delta$ in Theorem \ref{Thm1}.

\begin{description}
\item[(I)] In the case of $\Delta=(0, 0, 0, 0, 0, 0)$
\end{description}
\begin{eqnarray}
	T^{t+2}_{j+1} + T^t_j = \frac{1}{2} {}^t \!\mathbf{m} A \mathbf{m} + {}^t \!(\mathbf{z} + 2\mathbf{\Omega} - \mathbf{1})  \mathbf{m} + \frac{1}{2} {}^t \!\mathbf{m} A \mathbf{m} + {}^t \!\mathbf{z}  \mathbf{m} \nonumber\\
	= \frac{1}{2} {}^t \!\mathbf{m} A \mathbf{m} + {}^t \!(\mathbf{z} + \mathbf{\Omega} - \mathbf{1})  \mathbf{m} + \frac{1}{2} {}^t \!\mathbf{m} A \mathbf{m} + {}^t \!(\mathbf{z} + \mathbf{\Omega}) \mathbf{m} \nonumber\\
	= T^{t+1}_{j+1} + T^{t+1}_j \label{caseA1}
\end{eqnarray}

\begin{eqnarray}
	T^{t+2}_{j+1} + T^t_j = \frac{1}{2} {}^t \!\mathbf{m} A \mathbf{m} + {}^t \!(\mathbf{z} + 2\mathbf{\Omega})  \mathbf{m} + \frac{1}{2} {}^t \!\mathbf{m} A \mathbf{m} + {}^t \!(\mathbf{z} - \mathbf{1}) \mathbf{m} \nonumber\\
	= T^{t+2}_j + T^t_{j+1} > T^{t+2}_j + T^t_{j+1} - 1
\end{eqnarray}

\begin{description}
\item[(II)] In the case of $\Delta=(\mathbf{e}_i, 0, 0, 0, 0, 0)$
\end{description}
\begin{eqnarray}
	T^{t+2}_{j+1} + T^t_j \ge \frac{1}{2} {}^t \!(\mathbf{m} + \mathbf{e}_i) A (\mathbf{m} + \mathbf{e}_i) + {}^t \!(\mathbf{z} + 2\mathbf{\Omega} - \mathbf{1}) (\mathbf{m} + \mathbf{e}_i) + \frac{1}{2} {}^t \!\mathbf{m} A \mathbf{m} + {}^t \! \mathbf{z} \mathbf{m} \nonumber\\
	= \frac{1}{2} {}^t \!(\mathbf{m} + \mathbf{e}_i) A (\mathbf{m} + \mathbf{e}_i) + {}^t \!(\mathbf{z} + 2\mathbf{\Omega})  (\mathbf{m} + \mathbf{e}_i) + \frac{1}{2} {}^t \!\mathbf{m} A \mathbf{m} + {}^t \!(\mathbf{z} - \mathbf{1}) \mathbf{m} - 1 \nonumber\\
	= T^{t+2}_j + T^t_{j+1} - 1.
\end{eqnarray}

The proof for $T^{t+2}_{j+1} + T^t_j = T^{t+1}_{j+1} + T^{t+1}_j$ is the same as that for (\ref{caseA1}) by the definition of $T^{t+2}_{j+1}$.

\begin{description}
\item[(III)] In the case of $\Delta=(\mathbf{e}_i, \mathbf{e}_i, \mathbf{d}, \mathbf{d}, 0, 0)$
\end{description}
\begin{eqnarray}
	T^{t+2}_{j+1} + T^t_j = \frac{1}{2} {}^t \!(\mathbf{m} + \mathbf{e}_i) A (\mathbf{m} + \mathbf{e}_i) + {}^t \!(\mathbf{z} + 2\mathbf{\Omega} - \mathbf{1})  (\mathbf{m} + \mathbf{e}_i) + \frac{1}{2} {}^t \!\mathbf{m} A \mathbf{m} + {}^t \!\mathbf{z}  \mathbf{m} \nonumber\\
	= \frac{1}{2} {}^t \!(\mathbf{m} + \mathbf{e}_i) A (\mathbf{m} + \mathbf{e}_i) + {}^t \!(\mathbf{z} + 2\mathbf{\Omega})  (\mathbf{m} + \mathbf{e}_i) + \frac{1}{2} {}^t \!\mathbf{m} A \mathbf{m} + {}^t \!(\mathbf{z} - \mathbf{1}) \mathbf{m} - 1 \nonumber\\
	= T^{t+2}_j + T^t_{j+1} - 1 \label{caseA2}
\end{eqnarray}

\begin{eqnarray}
	T^{t+2}_{j+1} + T^t_j \ge \frac{1}{2} {}^t \!(\mathbf{m} + \mathbf{d}) A (\mathbf{m} + \mathbf{d}) + {}^t \!(\mathbf{z} + 2\mathbf{\Omega} - \mathbf{1})  (\mathbf{m} + \mathbf{d}) \nonumber\\
	+ \frac{1}{2} {}^t \!(\mathbf{m} + \mathbf{d}) A (\mathbf{m} + \mathbf{d}) + {}^t \!\mathbf{z} (\mathbf{m} + \mathbf{d}) \nonumber\\
	= \frac{1}{2} {}^t \!(\mathbf{m} + \mathbf{d}) A (\mathbf{m} + \mathbf{d}) + {}^t \!(\mathbf{z} + \mathbf{\Omega} - \mathbf{1}) (\mathbf{m} + \mathbf{d}) \nonumber\\
	+ \frac{1}{2} {}^t \!(\mathbf{m} + \mathbf{d}) A (\mathbf{m} + \mathbf{d}) + {}^t \!(\mathbf{z} + \mathbf{\Omega}) (\mathbf{m} + \mathbf{d}) \nonumber\\
	= T^{t+1}_{j+1} + T^{t+1}_j
\end{eqnarray}

\begin{description}
\item[(IV)] In the case of $\Delta=(\mathbf{e}_i, \mathbf{e}_i, \mathbf{d} + \mathbf{e}_k, \mathbf{d}, 0, 0)$
\end{description}
\begin{eqnarray}
	T^{t+2}_{j+1} + T^t_j \ge \frac{1}{2} {}^t \!(\mathbf{m} + \mathbf{d} + \mathbf{e}_k) A (\mathbf{m} + \mathbf{d} + \mathbf{e}_k) + {}^t \!(\mathbf{z} + \mathbf{\Omega}) (\mathbf{m} + \mathbf{d} + \mathbf{e}_k) \nonumber\\
	 + \frac{1}{2} {}^t \!(\mathbf{m} + \mathbf{d}) A (\mathbf{m} + \mathbf{d}) + {}^t \!(\mathbf{z} + \mathbf{\Omega} - \mathbf{1}) (\mathbf{m} + \mathbf{d}) + (\Omega_k - 1) \nonumber\\
	\ge T^{t+1}_{j+1} + T^{t+1}_j
\end{eqnarray}

The proof for $T^{t+2}_{j+1} + T^t_j = T^{t+2}_j + T^t_{j+1} - 1$ is the same as (\ref{caseA2}).

\begin{description}
\item[(V)] In the case of $\Delta=(\mathbf{e}_i, \mathbf{e}_i, \mathbf{d} + \mathbf{e}_k, \mathbf{d} + \mathbf{e}_k, 0, 0)$
\end{description}
The proof is completely the same as for (III).

\begin{description}
\item[(VI)] In the case of $\Delta=(\mathbf{e}_i, \mathbf{e}_i, \mathbf{e}_i, \mathbf{e}_i, \mathbf{e}_i, 0)$.
\end{description}
\begin{eqnarray}
	T^{t+2}_{j+1} + T^t_j = \frac{1}{2} {}^t \!(\mathbf{m} + \mathbf{e}_i) A (\mathbf{m} + \mathbf{e}_i) + {}^t \!(\mathbf{z} + 2\mathbf{\Omega})  (\mathbf{m} + \mathbf{e}_i) \nonumber\\
	 + \frac{1}{2} {}^t \!(\mathbf{m} + \mathbf{e}_i) A (\mathbf{m} + \mathbf{e}_i) + {}^t \!(\mathbf{z} - \mathbf{1}) (\mathbf{m} + \mathbf{e}_i) \nonumber\\
	\ge \frac{1}{2} {}^t \!(\mathbf{m} + \mathbf{e}_i) A (\mathbf{m} + \mathbf{e}_i) + {}^t \!(\mathbf{z} + 2\mathbf{\Omega})  (\mathbf{m} + \mathbf{e}_i) + \frac{1}{2} {}^t \!\mathbf{m} A \mathbf{m} + {}^t \!(\mathbf{z} - \mathbf{1}) \mathbf{m} \nonumber\\
	= T^{t+2}_j + T^t_{j+1} > T^{t+2}_j + T^t_{j+1} - 1
\end{eqnarray}

The proof for $T^{t+2}_{j+1} + T^t_j = T^{t+1}_{j+1} + T^{t+1}_j$ is the same as (\ref{caseA1}).

\begin{description}
\item[(VII)] In the case of $\Delta=(2\mathbf{e}_i, \mathbf{e}_i, \mathbf{e}_i, \mathbf{e}_i, \mathbf{e}_i, 0)$.
\end{description}

The proof for $T^{t+2}_{j+1} + T^t_j = T^{t+1}_{j+1} + T^{t+1}_j$ is the same as (\ref{caseA1}) and by virtue of (\ref{sqidentity}), one has
\begin{eqnarray}
	T^{t+2}_{j+1} + T^t_j = \frac{1}{2} {}^t \!(\mathbf{m} + \mathbf{e}_i) A (\mathbf{m} + \mathbf{e}_i) + {}^t \!(\mathbf{z} + 2\mathbf{\Omega} - \mathbf{1})  (\mathbf{m} + \mathbf{e}_i) \nonumber\\ 
	 + \frac{1}{2} {}^t \!(\mathbf{m} + \mathbf{e}_i) A (\mathbf{m} + \mathbf{e}_i) + {}^t \!\mathbf{z} (\mathbf{m} + \mathbf{e}_i) \nonumber\\
	= \frac{1}{2} {}^t \!(\mathbf{m} + 2 \mathbf{e}_i) A (\mathbf{m} + 2 \mathbf{e}_i) + {}^t \!(\mathbf{z} + 2\mathbf{\Omega})  (\mathbf{m} + 2 \mathbf{e}_i) \nonumber\\ 
	+ \frac{1}{2} {}^t \!\mathbf{m} A \mathbf{m} + {}^t \!(\mathbf{z} - \mathbf{1}) \mathbf{m} - 1 + (A)_{i, i} - 2 \Omega_i.
\end{eqnarray}
Due to relation (\ref{condL}), we obtain
\begin{equation}
	T^{t+2}_{j+1} + T^t_j > T^{t+2}_j + T^t_{j+1} - 1.
\end{equation}

\begin{description}
\item[(VIII)] In the case of $\Delta=(\mathbf{e}_i + \mathbf{e}_l, \mathbf{e}_i, \mathbf{e}_i, \mathbf{e}_i, \mathbf{e}_i, 0)$.
\end{description}

The proof for $T^{t+2}_{j+1} + T^t_j = T^{t+1}_{j+1} + T^{t+1}_j$ is the same as (\ref{caseA1}) and by virtue of (\ref{sqidentity}), one has
\begin{eqnarray}
	T^{t+2}_{j+1} + T^t_j = \frac{1}{2} {}^t \!(\mathbf{m} + \mathbf{e}_i) A (\mathbf{m} + \mathbf{e}_i) + {}^t \!(\mathbf{z} + 2\mathbf{\Omega} - \mathbf{1})  (\mathbf{m} + \mathbf{e}_i) \nonumber\\ 
	 + \frac{1}{2} {}^t \!(\mathbf{m} + \mathbf{e}_i) A (\mathbf{m} + \mathbf{e}_i) + {}^t \!\mathbf{z} (\mathbf{m} + \mathbf{e}_i) \nonumber\\
	= \frac{1}{2} {}^t \!(\mathbf{m} + 2 \mathbf{e}_i) A (\mathbf{m} + 2 \mathbf{e}_i) + {}^t \!(\mathbf{z} + 2\mathbf{\Omega})  (\mathbf{m} + 2 \mathbf{e}_i) \nonumber\\ 
	+ \frac{1}{2} {}^t \!\mathbf{m} A \mathbf{m} + {}^t \!(\mathbf{z} - \mathbf{1}) \mathbf{m} - 1 + (A)_{i, i} - 2 \Omega_i.
\end{eqnarray}
Due to relation (\ref{condL}), we obtain
\begin{equation}
	T^{t+2}_{j+1} + T^t_j > T^{t+2}_j + T^t_{j+1} - 1.
\end{equation}

\begin{description}
\item[(IX)] In the case of $\Delta=(\mathbf{e}_i + \mathbf{e}_k, \mathbf{e}_i, \mathbf{d} + \mathbf{e}_{k'}, \mathbf{d} + \mathbf{e}_{k'}, \mathbf{e}_k, 0)$
\end{description}
By virtue of (\ref{sqidentity}) and $i>k$, one has $(A)_{i, k} = - 2 \Omega_k = -2 {}^t \! \mathbf{\Omega} \mathbf{e}_k$. Then,
\begin{eqnarray}
	T^{t+2}_{j+1} + T^t_j = \frac{1}{2} {}^t \!(\mathbf{m} + \mathbf{e}_i) A (\mathbf{m} + \mathbf{e}_i) + {}^t \!(\mathbf{z} + 2\mathbf{\Omega} - \mathbf{1})  (\mathbf{m} + \mathbf{e}_i) \nonumber\\ 
	 + \frac{1}{2} {}^t \!(\mathbf{m} + \mathbf{e}_k) A (\mathbf{m} + \mathbf{e}_k) + {}^t \!\mathbf{z} (\mathbf{m} + \mathbf{e}_k) \nonumber\\
	= \frac{1}{2} {}^t \!(\mathbf{m} + \mathbf{e}_i + \mathbf{e}_k) A (\mathbf{m} + \mathbf{e}_i + \mathbf{e}_k) \nonumber\\ 
	 + {}^t \!(\mathbf{z} + 2\mathbf{\Omega})  (\mathbf{m} + \mathbf{e}_i + \mathbf{e}_k) + \frac{1}{2} {}^t \!\mathbf{m} A \mathbf{m} + {}^t \!(\mathbf{z} - \mathbf{1}) \mathbf{m} - 1 \nonumber\\
	= T^{t+2}_j + T^t_{j+1} - 1.
\end{eqnarray}
We also obtain $T^{t+2}_{j+1} + T^t_j \ge T^{t+1}_{j+1} + T^{t+1}_j$ by the same proof in the latter part of (III). \hfill \opensquare
\end{Proof}

\section{Examples}

In this section, we study several aspects of the behaviour of the functions we proposed. We employ the dependent variable $U^t_j$ defined in (\ref{defu}) in the  plots, because this variable is best suited for observing the behaviour we are interested in. In the following figures we depict the $j$-lattice by a row of boxes, containing a ball when $U^t_j$ is equal to $1$, and empty whenever $U^t_j=0$.

\subsection{Infinite Domain}

\begin{figure}
\begin{center}
\includegraphics[scale=0.66]{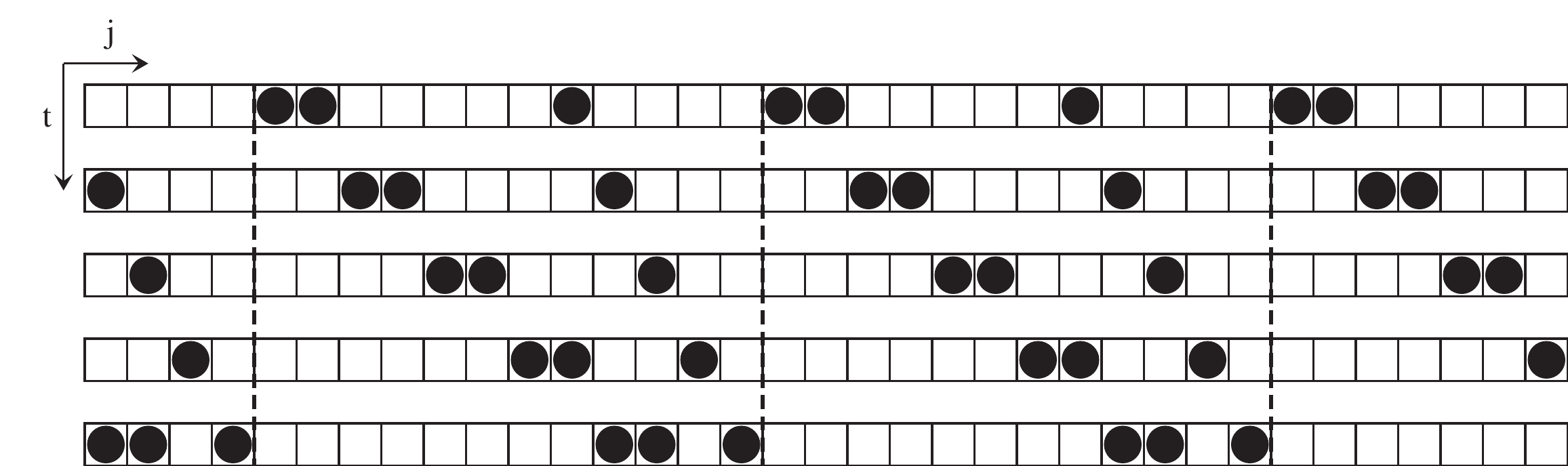}
\caption{$U^t_j$ for $N=2, \mathcal{D}=\mathbb{Z}^2, 0 \le t \le 4, 0 \le j \le 34, \Omega_1=1, \Omega_2=2, C_1=22, C_2=17, L_1=12, L_2=12$. The area between dashed lines is one period.}\label{figure3}
\end{center}
\end{figure}

In this subsection, we treat the case where ${\mathcal D} = \mathbb{Z}^N$ and assume that all parameters ($\Omega_i$, $C_i$, $L_i$) are integers. Before arguing the general case, we first consider a special case where the solutions are reduced to well-known ones. Let $L_1 = \ldots = L_N = L$, and we obtain the identity:
\begin{equation}
	f(\mathbf{z}^t_{j+L}; \mathbf{m} - \mathbf{1}) = f(\mathbf{z}^t_j; \mathbf{m}) - {}^t \! (\mathbf{z}^t_j - \frac{1}{2} L \mathbf{1}) \mathbf{1} \label{identity}
\end{equation}
because of $A \mathbf{1} = - L \mathbf{1}$. Due to this identity, if $\mathbf{m}$ yields the maximum of $f(\mathbf{z}; \cdot)$, $\mathbf{m} - \mathbf{1}$ also yields that of $f(\mathbf{z} - \mathbf{1}; \cdot)$. In particular, by substituting $\mathbf{z} = \mathbf{z}^t_j$, one has:
\begin{eqnarray}
	\mathbf{m}^t_{j+L} = \mathbf{m}^t_j - \mathbf{1} \\
	T^t_{j+L} = T^t_j - \sum_{i=1}^N (t \Omega_i - j + C_i) + \frac{1}{2} N L. \label{pseudoperiodicity}
\end{eqnarray}
By virtue of (\ref{pseudoperiodicity}), we obtain the relationship $U^t_{j+L} = U^t_j$, where $T^t_j$ expresses a state of the pBBS. Indeed, this is known as a standard form of solutions for the pBBS. We note that the condition (\ref{condL}) simplifies to $2 \sum_{i=1}^N \Omega_i < L$, which is a famous requirement in the analysis of pBBS. Figure \ref{figure3} depicts an example of such solutions.

\begin{figure}
\begin{center}
\includegraphics[scale=0.66]{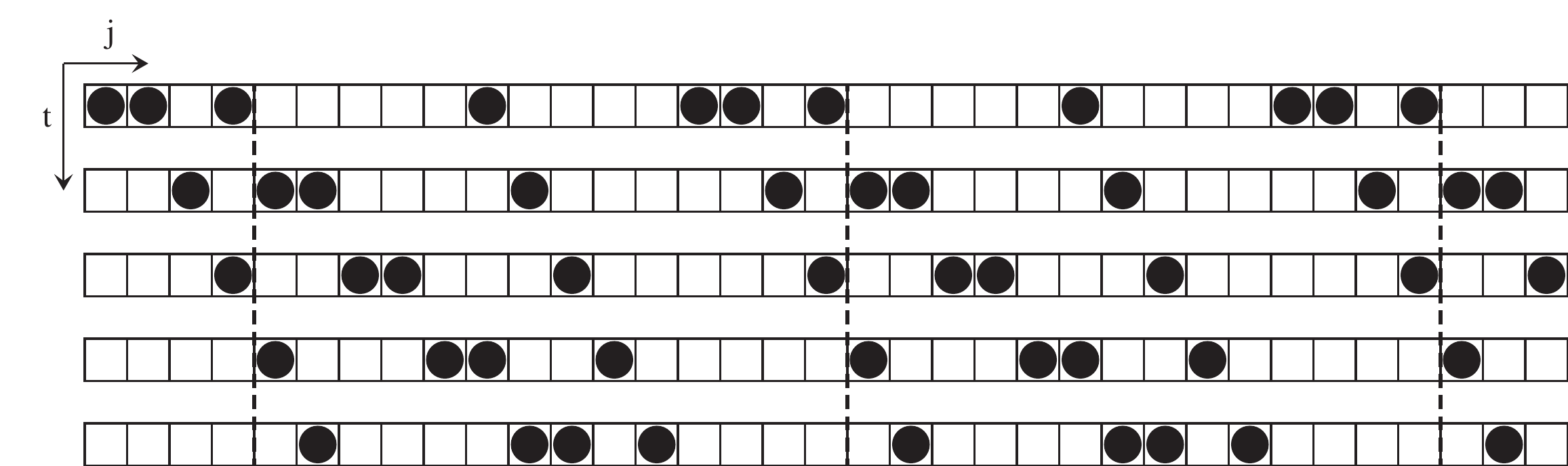}
\caption{By plotting $U^t_j$ for $N=2, \mathcal{D}=\mathbb{Z}^2, 0 \le t \le 4, 0 \le j \le 34, \Omega_1=1, \Omega_2=2, C_1=22, C_2=17, L_1=8, L_2=12$. The area between dashed lines is one period.}\label{figure4}
\end{center}
\end{figure}

We now consider a generalizations of (\ref{identity})--(\ref{pseudoperiodicity}). If we can find $K>0$ and $L_1, \ldots, L_N$ satisfying
\begin{equation} \label{identity1}
	A \mathbf{n} = - K \mathbf{1}
\end{equation}
for given $\mathbf{n} \in \mathbb{Z}_{>0}^N$, we obtain the relationship
\begin{eqnarray}
	\mathbf{m}^t_{j+K} = \mathbf{m}^t_j - \mathbf{n} \\
	U^t_{j+K} = U^t_j,
\end{eqnarray}
by employing the similar argument. In this case, $T^t_j$ also expresses a state of the pBBS. The main difference from the standard form is that the each block of balls parametrized with ($\Omega_i$, $C_i$) emerges $(\mathbf{n})_i$ times in a single period. For such cases, we should take $N$ to be the number of apparent blocks when we employ the standard form. However, in our representation, we may employ smaller $N$. Therefore, the function is in fact a ``compressed" representation for this system. Figure \ref{figure4} depicts an example of such solutions.

\begin{figure}\
\includegraphics[scale=0.66]{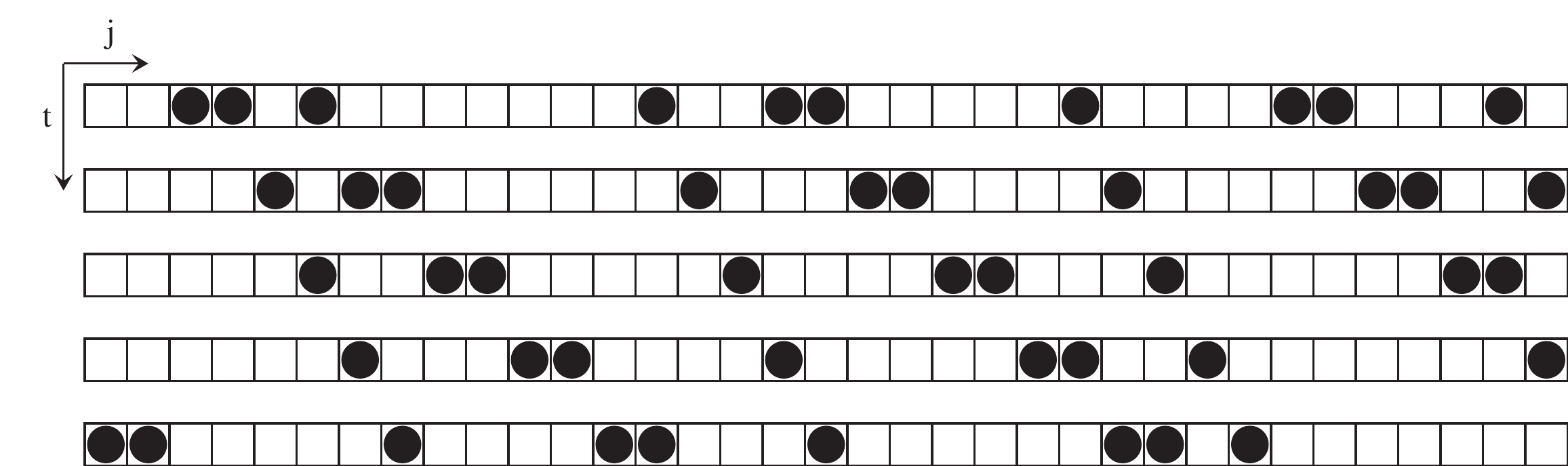}
\caption{By plotting $U^t_j$ for $N=2, \mathcal{D}=\mathbb{Z}^2, 0 \le t \le 4, 0 \le j \le 34, \Omega_1=1, \Omega_2=2, C_1=22, C_2=17, L_1=10, L_2=12$.}\label{figure2}
\end{figure}

Finally we consider the case where $L_1, \ldots, L_N$ are given, as shown in Figure \ref{figure2}. We observe that each block of balls parametrized by ($\Omega_i$, $C_i$) has its own ``pattern", depending on $L_i$. To discover its global behaviour, we consider the equations:
\begin{equation}
	(A \mathbf{n})_1 = \ldots = (A \mathbf{n})_N.
\end{equation}
Since all coefficients of $A$ are integers, by the assumption, we can obtain $\mathbf{n} \in \mathbb{Z}^N$ and $K > 0$ satisfying the relationship: $A \mathbf{n} = - K \mathbf{1}$. Therefore, by virtue of the same arguments as above, solutions are always periodic, for general parameters. In fact, the solution depicted in Figure \ref{figure2} has a period of $38$. We note that it is hard to predict the period, directly from parameters. All solutions depicted in Figures \ref{figure3} -- \ref{figure2} take the same parameters ($\Omega_i$ $C_i$) and $L_2$. However, we observe that just a little change for $L_1$ provokes a large difference for the periods.

\subsection{Finite Domain}

\begin{figure}
\begin{center}
\includegraphics[scale=0.66]{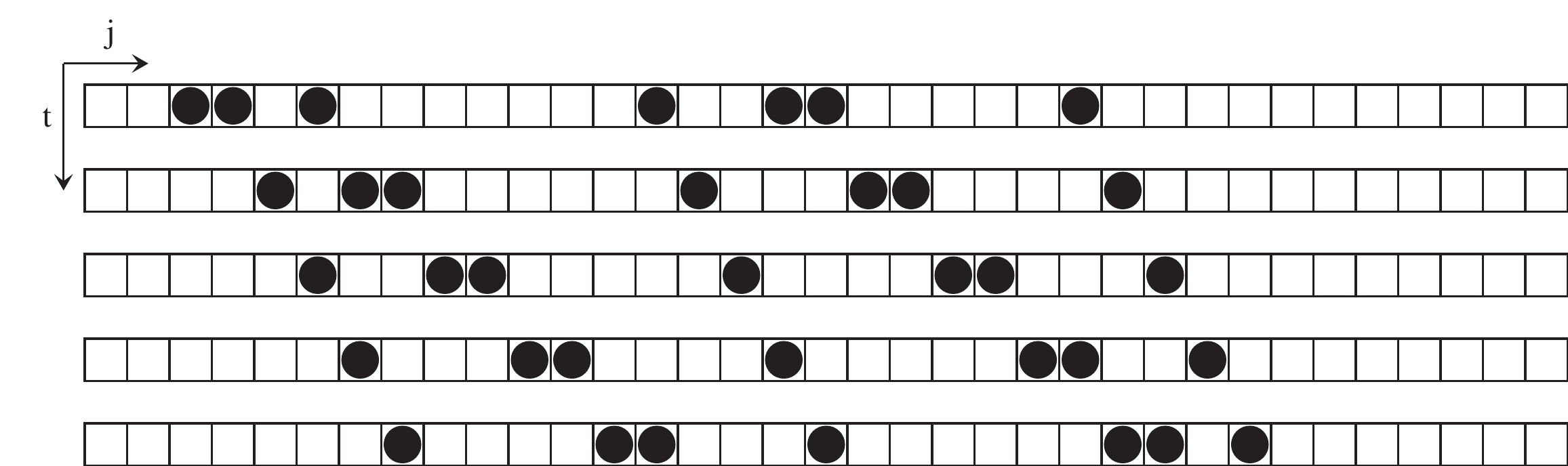}
\caption{By plotting $U^t_j$ for $N=2, \mathcal{D}=[-1,2] \times [-1, 1], 0 \le t \le 4, 0 \le j \le 34, \Omega_1=1, \Omega_2=2, C_1=22, C_2=17, L_1=8, L_2=12$.}\label{figure5}
\end{center}
\end{figure}

We next consider the case where all ${\mathcal D}_i$ are finite. Figure \ref{figure5}. depicts an example of such a case. Blocks of balls emerge several times but disappear in distant sites. Therefore, this solution satisfies the boundary condition $U^t_j = 0$ for $|j| \gg 1$ and solves (\ref{evBBS}) for $j_0 = -\infty$, which is the time evolution of the standard BBS. Especially in the case $D_i = [0, 1]$ (for all $i$), each block appears only one time and the solutions express the well known soliton solutions for this system. This type of solution is also a compressed representation for the state with regularly-positioned blocks of balls. However, this compressed solution is easily rewritten from the non-compressed (well-known) one, i.e., it is not a new solution. 

Of course, we can consider the case where the $D_i$ are semi-infinite. For such cases, the corresponding block appears in a pattern for sufficiently large $j$ but never for sufficiently small $j$ (or vice versa). We can also take ${\mathcal D}$ to be a product of finite interval and infinite one, which expresses the state where some blocks appear infinitely many times but other blocks appear only finitely many times.

Finally, we note that limiting the domain corresponds to truncating summations for discrete systems and we stress that the truncated solutions of discrete integrable equations cease being solutions.

\section{Concluding Remarks}

In this paper, we have discussed properties for a class of multi-variable quadratic functions and we have proven that these functions solve the ultradiscrete KdV equation. We have also proposed a new type of solutions by restricting parameters, for example, multi-periodic solutions. 

\bigskip

In our previous papers \cite{Nakata2009}, \cite{Nakata2010}, we proposed a recursive representation of soliton solutions for the ultradiscrete soliton equations including the ultradiscrete KdV equation. This representation can be considered as a transformation from a soliton solution to another one, that is, a transformation between two states in the same dynamics---the standard BBS, with infinite and open boundary condition.
However, the recursive representation (\ref{vertexop}) cannot be regarded as transformation between two ``standard" forms of solutions to pBBS, because of the shift of the parameter $L_i$ between $T^{(N), t}_j$ and $\tilde{T}^{(N-1), t}_j$. In other words, when we obtain a standard form by applying the recursive form (\ref{vertexop}) repeatedly from the vacuum solution $T^{(0), t}_j \equiv 0$, the intermediate states are not standard forms except for some special cases (for example when all $\Omega_i$ are equal).
It should be noted that relation (\ref{vertexop}) is an extension of a recursive representation of soliton solutions. When restricting $D_N = [0, 1]$, we can omit the term of $m_N^2$ in (\ref{vertexop}) by replacing $C_N \to C_N + L_N - 2 \sum_{i=1}^{N-1} \Omega_i$ because of $m_N = m_N^2$ for $m_N = 0, 1$. The representation presented here is suitable for only analyzing the standard BBS.

\bigskip

As seen in the proof of Theorem \ref{mainThm}, it is important to consider the state of $\mathbf{m}^t_j$, which corresponds to each state of the BBS. Especially, the case (IX) corresponds to the interaction of solitons. In \cite{Nakata2011}, it was sufficient to consider only two cases, interacting or not, to describe the dynamics of the ultradiscrete Toda molecule equation. However, for the ultradiscrete KdV equation,we also have to consider an additional case --- ``injecting" balls. The reason for this complexity results from the introduction of coordinates $j$ for boxes to represent the Box and Ball dynamics.

It is also known that the BBS has waves called ``backgrounds" which can take various value and travel at speed $1$ and arbitrary initial states consist of solitons and backgrounds. Cauchy problems can be exactly solved by virtue of an ultradiscrete analogue of the inverse scattering method \cite{WilloxNakataSatsumaRamaniGrammaticos}. However, our approach in this paper cannot be applied to states that include backgrounds, as it depends strongly on good combinatorial properties of the solitons.

\bigskip

We also note that the discussion in the previous sections becomes much easier in the case $D_i = [0, 1]$ (for all $i$) because we consider only limited cases. Specifically, we do not need to employ Proposition \ref{negative}, or most of Theorem \ref{relt}. Therefore, we can present an induction free discussion, such as proposed in \cite{Nakata2011}.

\bigskip

We believe that discrete convexity plays a very important role for the $\tau$-functions of the ultradiscrete systems. In this paper however, we only used  convexity for some propositions which are not so essential. The quadratic function (\ref{deff}) is certainly convex and has good combinatorial properties. However, we cannot obtain good properties of the solution $T^{(N), t}_j$ itself. We also note that the independent variables $t$ and $j$ in (\ref{deftau}) can be extended to real values, but $\mathbf{m}$ in (\ref{deff}) can not. The discreteness for $\mathbf{m}$ is considered to be essential. It is expected that a direct relationship between these notions and fundamental properties of integrable systems such as the Pl\"ucker relations, and can be expressed in the language of discrete convex analysis.

\section*{Acknowledgment}
The author would like to thanks  Professors T. Tokihiro and R. Willox for helpful comments.


\section*{References}

\begin{thebibliography}{10}

\bibitem{Hirota1977-1}
R.~Hirota.
\newblock {Nonlinear Partial Difference Equations I; A Difference Analogue of
  the Korteweg-de Vries Equation}.
\newblock {\em J. Phys. Soc. Jpn.}, 43:1424--1433, 1977.

\bibitem{TakahashiTokihiroMatsukidairaSatsuma}
T.~Tokihiro, D.~Takahashi, J.~Matsukidaira, and J.~Satsuma.
\newblock {From Soliton Equations to Integrable Cellular Automata through a
  Limiting Procedure}.
\newblock {\em Phys. Rev. Lett.}, 76:3247--3250, 1996.

\bibitem{TakahashiSatsuma}
D.~Takahashi and J.~Satsuma.
\newblock A soliton cellular automaton.
\newblock {\em J. Phys. Soc. Jpn.}, 59:3514--3519, 1990.

\bibitem{Nakata2009}
Y.~Nakata.
\newblock {Vertex operator for the ultradiscrete KdV equation}.
\newblock {\em J. Phys. A: Math. Theor.}, 42:412001 (6pp), 2009.

\bibitem{Nakata2010}
Y.~Nakata.
\newblock {Vertex operator for the non-autonomous ultradiscrete KP equation}.
\newblock {\em J. Phys. A: Math. Theor.}, 43:195201 (8pp), 2010.

\bibitem{TakahashiHirota}
D.~Takahashi and R.~Hirota.
\newblock Ultradiscrete soliton solution of permanent type.
\newblock {\em J. Phys. Soc. Jpn.}, 76:104007, 2007.

\bibitem{NagaiTakahashi2010}
H.~Nagai and D.~Takahashi.
\newblock {Bilinear equations and Backlund transformation for a generalized
  ultradiscrete soliton solution}.
\newblock {\em J. Phys. A: Math. Theor.}, 43:375202 (13pp), 2010.

\bibitem{NagaiTakahashi2011}
H.~Nagai and D.~Takahashi.
\newblock {Ultradiscrete Pl\"ucker Relation Specialized for Soliton Solutions}.
\newblock {\em J. Phys. A: Math. Theor.}, 44:095202 (18pp), 2011.

\bibitem{TakagakiKamioka}
T.~Takagaki and S.~Kamioka.
\newblock {\em Proceedings of RIAM, Kyushu University (JAPANESE)}, 2011.

\bibitem{NoumiYamada}
M.~Noumi and Y.~Yamada.
\newblock {Tropical Robinson-Schensted-Knuth correspondence and birational Weyl
  group actions}.
\newblock Technical Report~40, Adv. Stud. Pure Math., 2004.

\bibitem{Nakata2011}
Y.~Nakata.
\newblock {Solutions to the ultradiscrete Toda molecule equation expressed as
  minimum weight flows of planar graphs}.
\newblock {\em J. Phys. A: Math. Theor.}, 44:295204 (15pp), 2011.

\bibitem{KimijimaTokihiro}
T.~Kimijima and T.~Tokihiro.
\newblock Initial-value problem of the discrete periodic toda equation and its
  ultradiscretization.
\newblock {\em Inverse Problems}, 18:1705--1732, 2002.

\bibitem{InoueTakenawa}
R.~Inoue and T.~Takenawa.
\newblock {Tropical spectral curves and integrable cellular automata}.
\newblock {\em Int. Math. Res. Not. IMRN}, (9):Art ID. rnn019, 27pp., 2008.

\bibitem{Iwao2009}
S.~Iwao.
\newblock {Integration over Tropical Plane Curves and Ultradiscretization}.
\newblock {\em Int. Math. Res. Not. IMRN}, 2010(1):112--148, 2009.

\bibitem{HatayamaHikamiInoueKunibaTakagi}
G.~Hatayama, K.~Hikami, R.~Inoue, A.~Kuniba, and T.~Takagi.
\newblock {The $A^{(1)}_{M}$ automata related to crystals of symmetric
  tensors}.
\newblock {\em J. Math. Phys.}, 42:274--308, 2001.

\bibitem{Murota2003}
K.~Murota.
\newblock {\em Discrete Convex Analysis}, volume~10.
\newblock Society for Industrial and Applied Mathematics, 2003.

\bibitem{WilloxNakataSatsumaRamaniGrammaticos}
R.~Willox, Y.~Nakata, J.~Satsuma, A.~Ramani, and B.~Grammaticos.
\newblock {Solving the ultradiscrete KdV equation}.
\newblock {\em J. Phys. A: Math. Theor.}, 43:482003 (7pp), 2010.

\end{thebibliography}
\end{document}